\documentclass[aps,prd,preprint,superscriptaddress,showpacs,floatfix,nobibnotes]{revtex4-1}

\usepackage{latexsym}
\usepackage{amsmath}
\usepackage{amssymb}
\usepackage{graphicx}
\usepackage{setspace}

\usepackage{placeins}

\usepackage{bm}
\usepackage{epsfig}
\usepackage{subfigure}
\newcommand{\bea}{\begin{eqnarray}}
\newcommand{\eea}{\end{eqnarray}}

\begin{document}

\title{Renormalization of the 4PI effective action using the functional renormalization group}

\author{M.E. Carrington}
\email[]{carrington@brandonu.ca} 
\affiliation{Department of Physics, Brandon University, Brandon, Manitoba, R7A 6A9 Canada}\affiliation{Winnipeg Institute for Theoretical Physics, Winnipeg, Manitoba}
 
\author{S.A. Friesen}
\email[]{friesenseth@gmail.com} 
\affiliation{Department of Physics, Brandon University, Brandon, Manitoba, R7A 6A9 Canada}

\author{C.D. Phillips}
\email[]{christopherdphillips7@gmail.com} 
\affiliation{Department of Physics, Brandon University, Brandon, Manitoba, R7A 6A9 Canada}

\author{D. Pickering}
\email[]{pickering@brandonu.ca} 
\affiliation{Department of Mathematics, Brandon University, Brandon, Manitoba, R7A 6A9 Canada}

\date{\today}

\begin{abstract}

Techniques based on $n$-particle irreducible effective actions can be used to study systems where perturbation theory does not apply. The main advantage, relative to other non-perturbative continuum methods, is that the hierarchy of integral equations that must be solved truncates at the level of the action, and no additional approximations are needed. 
The main problem with the method is renormalization, which until now could only be done at the lowest ($n$=2) level. 
In this paper we show  how to obtain renormalized results from an $n$-particle irreducible effective action at any order. 
We consider a symmetric scalar theory with quartic coupling in four dimensions and show that the 4 loop 4-particle-irreducible calculation can be renormalized using a renormalization group method. 
The calculation involves one bare mass and one bare coupling constant which are introduced at the level of the Lagrangian, and  cannot be done using any known method by introducing counterterms. 
 
\end{abstract}


\normalsize
\maketitle

\normalsize

\section{Introduction}
\label{section-introduction}

There are many interesting systems that are governed by non-perturbative physics. 
Quantum chromodynamics (QCD) is non-perturbative except at very high energy scales, and quark gluon plasma has been studied extensively for over 20 years as a physical system which could give access to fundamental properties of QCD. 
Another important example is three dimensional quantum electrodynamics (QED), which is physically relevant in the context of condensed matter physics. There has been an explosion of recent interest in QED$_{2+1}$ with the discovery of graphene, in light of its importance in technological applications. 
The general importance of the study of non-perturbative field theories
is evidenced by the amount of work has been invested in the development of several different theoretical methods including numerical lattice calculations, the AdS/CFT correspondence, various formulations of the renormalization group, and Schwinger-Dyson equations. In this work we study another method which has some distinct advantages: the $n$-particle-irreducible ($n$PI) effective action. 
The method was originally developed in the context of non-relativistic statistical mechanics \cite{Yang, Luttinger, Martin}. 
In its modern form, the $n$PI action is written as functional of dressed vertex functions which are determined self-consistently using the variational principle \cite{Jackiw1974,Norton1975}. 
One major advantage of the $n$PI method is that it provides a systematic expansion for which the truncation occurs at the level of the action. In contrast, the Schwinger-Dyson integral equations give an infinite coupled heirarchy which must be truncated by introducing some extra approximation. 
A major disadvantage of $n$PI methods is a violation of gauge invariance \cite{Smit2003,Zaraket2004}. 
A procedure to minimize gauge dependence has been proposed \cite{sym-improvement}, and some issues with applying the tehcnique are discussed in \cite{Marko20156,Whittingham2016,Whittingham2017}.

2PI effective actions have been used for almost 20 years to study the thermodynamics of quantum fields \cite{Blaizot1999,Berges2005a,meggison,sohrabi,phillips-crete}, transport coefficients \cite{Aarts2004,Carrington2006,Carrington-transport-3pi,carrington-transport-4pi}, and non-equilibrium quantum dynamics \cite{berges-cox,berges-aarts,berges-baier,berges-non,berges-aarts2,tranberg-smit,tranberg-aarts,tranberg-laurie}.
On the other hand, while higher order effective actions have been derived using several different methods  \cite{berges-hierarchy,Carrington2004,Guo2011,Guo2012}, very little progress has been made in solving the resulting variational equations. 
%
We comment that although we could try to ignore vertex corrections and improve previous 2PI calculations by increasing the order of the truncation (usually the loop order), it is known that $n$PI formulations with $n>2$ are necessary in some situations.  
For a symmetric scalar $\phi^4$ theory, it has been shown numerically that 4PI vertex corrections are important in 3 dimensions \cite{mikula}.
For the same theory, working in four dimensions, it has been shown  that the 2PI approximation breaks down at large coupling - in the sense that successive orders in the loop expansion give large corrections \cite{meggison,phillips-crete}. 
In addition, it is known that leading order transport coefficients in QED and QCD cannot be obtained using a 2PI formulation \cite{Carrington-transport-3pi}.
Finally, there are general arguments that an $L$ loop calculation in the $n$PI formalism should be done with $L=n$.
Firstly, the $n$-loop $n$PI calculation is complete, in the sense that increasing the order of the approximation (the number of variational vertices that are included) without increasing the loop order of the truncation does not change the effective action \cite{berges-hierarchy}. 
Secondly, in gauge theories, the $n$ loop $n$PI effective action respects gauge invariance to the order of the truncation \cite{Smit2003,Zaraket2004} and one therefore expects that, for example, a 3 loop 2PI calculation will have stronger gauge dependence than a 3 loop 3PI one.

%
In this paper we are concerned with the renormalization of $n$PI actions, which is a fundamental problem that must be addressed before any calculations beyond the leading 2PI level of approximation can be attempted. 
%
%
The basic problem is that the self-consistent sets of integral equations that one must solve are plagued by ultraviolet divergences
and, beyond the 2PI level \cite{vanHees2002,Blaizot2003,Serreau2005}, 
a procedure for constructing the counterterms needed to eliminate the
corresponding divergences is unknown. 
The problem of renormalization is particularly complicated at finite temperature. 
One expects on general grounds that ultraviolet divergences
should be unaffected by the temperature, but temperature dependent divergences can appear in self-consistent approximations, and would cast doubt on the possibility of extracting physical quantities from the method. 

In this paper we develop and implement a functional renormalization group (FRG) method to renormalize the 4PI theory at the 4 loop level (for some related works see     
\cite{Blaizot-2PIa,Blaizot-2PIb,Carrington-BS,pulver, sohrabi,Dupuis1,Dupuis2,Katanin}). No counterterms are introduced, and all divergences are absorbed into the bare parameters of the Lagrangian, the structure of which is fixed and completely independent of the order of the approximation. 
We work (so far) with a symmetric scalar theory, in order to avoid the complications associated with the Lorentz and Dirac structures of fields in gauge theories.
%

This paper is organized as follows. In section \ref{section-preliminaries} we present our notation and the setup of the calculation. 
In section \ref{section-method} we describe our method and derive the flow equations that we will solve. In section \ref{section-results} we give some details of our numerical procedure and present our results. Conclusions are given in section \ref{section-conclusions}.

\section{Preliminaries}
\label{section-preliminaries}

Using standard notation we suppress the arguments that denote the space-time dependence of functions. For example, the term in the action that is quadratic in the field becomes
\bea
\frac{i}{2}\int d^4 x\,d^4 y\,\varphi(x)G_{\rm no\cdot int}^{-1}(x-y)\varphi(y) ~~\longrightarrow~~\frac{i}{2}\varphi\, G_{\rm no\cdot int}^{-1}\varphi\,.
\eea
In our notation $G_{\rm no\cdot int}$ is the bare propagator and the classical action is 
\bea
\label{action}
&& S[\varphi] =\frac{i}{2}\varphi \,G_{\rm no\cdot int}^{-1}\varphi -\frac{i}{4!}\lambda\varphi^{4}\,,~~
iG_{\rm no\cdot int}^{-1} = -(\Box + m^2)\,.
\eea
We use a scaled version of the physical coupling constant
($\lambda_{\,{\rm phys}} = i\lambda$). The extra factor of $i$ is introduced for notational convenience  and will be removed when rotating to Euclidean space to do numerical calculations. 

\begin{figure}[h]
\begin{center}
\includegraphics[width=8cm]{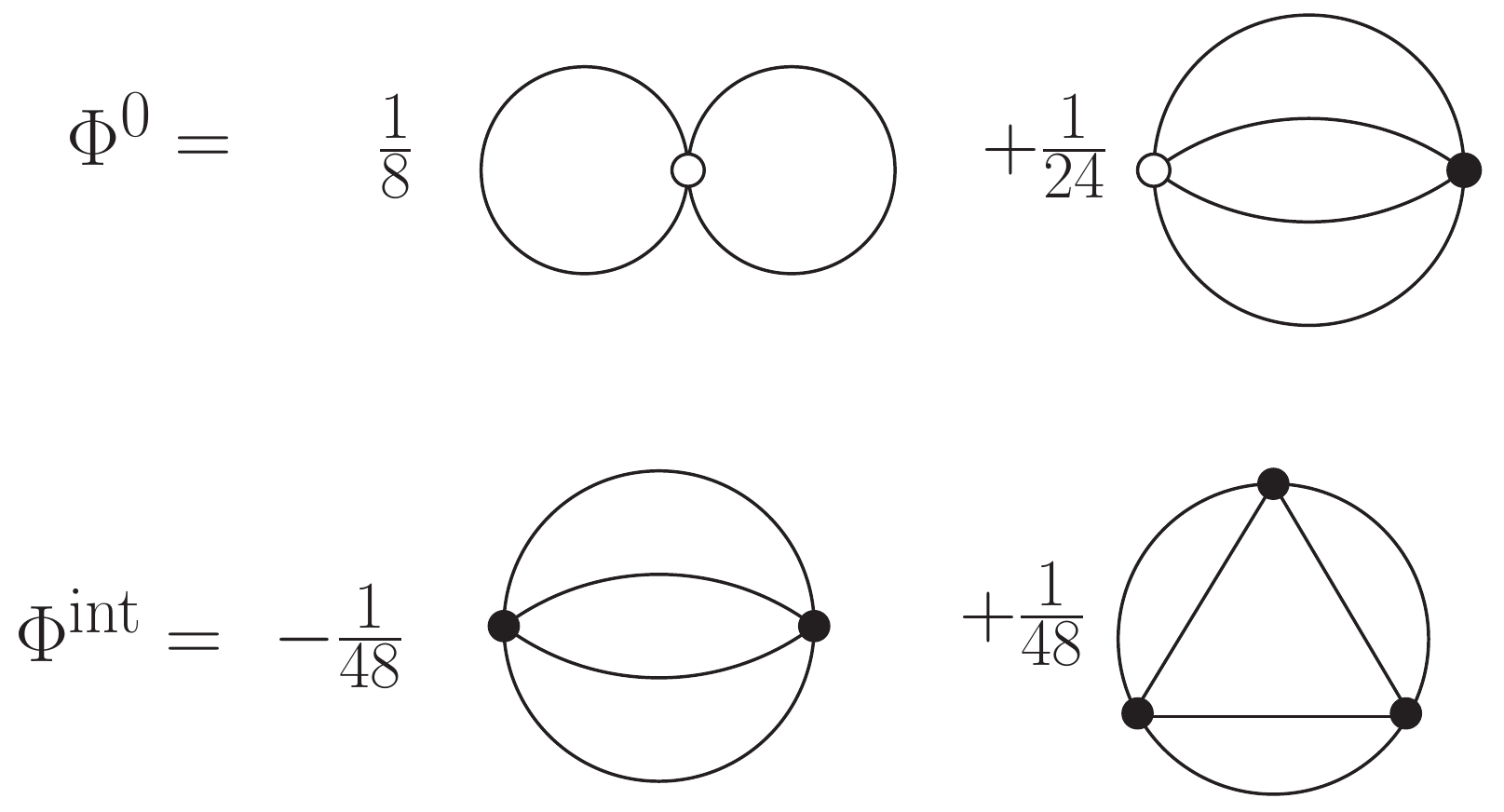}
\end{center}
\caption{The effective action for the symmetric theory to 4 loop order.  \label{phi-fig}}
\end{figure}

The $n$PI effective action is obtained by taking the $n$th Legendre transform of the generating
functional which is constructed by coupling the field to $n$ source terms. We will use $G$ for a self-consistent propagator and $V$ for a self-consistent vertex. 
The result for the 4 loop 4PI effective action in the symmetric theory has the form \cite{Guo2011,Guo2012}
\bea
\label{gammaGen}
\Gamma[D,V] = 
    \frac{i}{2} {\rm Tr} \,{\rm Ln}G^{-1}  +
\frac{i}{2} {\rm Tr}\left(G_{\rm no\cdot int}^{-1} G\right) -i\Phi^0[G,V]-i\Phi^{\rm int}[G,V]~~+~~{\rm const} 
\eea
where the terms $\Phi^0[G,V]$ and $\Phi^{\rm int}[G,V]$ contain all contributions to the effective action with two or more loops. All bare vertices are in the piece $\Phi^0[G,V]$.
The diagrams in $\Phi^0[G,V]$ and $\Phi^{\rm int}[G,V]$ at the four loop level are shown in Fig. \ref{phi-fig}. 

We now discuss the functional renormalization group method. 
The basis of the method is that we add to the action in (\ref{action}) a non-local regulator term 
\bea
\label{action-RG}
S_{\kappa}[\varphi]=S[\varphi]+\Delta S_{\kappa}[\varphi]\,,~~~
\Delta  S_{\kappa}[\varphi] = -\frac{1}{2}\hat R_{\kappa}\varphi^2\,. 
\eea
The parameter $\kappa$ has dimensions of momentum and its role is to divide the full momentum range into two regions which lie above and below the scale $\kappa$. 
The key point is that the regulator function is chosen so that $\lim_{Q\ll \kappa}\hat R_{\kappa}(Q)\sim \kappa^{2}$ and $\lim_{Q\geq\kappa}\hat R_{\kappa}(Q)\rightarrow 0$. 
The result is that for $Q\ll \kappa$ the regulator plays the role of a large mass term 
which suppresses quantum fluctuations with wavelengths $1/Q\gg
1/\kappa$, and in the opposite limit when $Q\gg\kappa$ the regulator goes to zero and fluctuations with wavelengths $1/Q\ll
1/\kappa$ are not affected.
As we will explain below, the basis of the method is to chose an initial value for $\kappa$ which is much greater than any other momentum scale in the problem, so that the theory initially behaves classically, and then slowly lower $\kappa$ to zero in such a way that a finite quantum theory is produced.

It is easy to see that if we include the FRG regulator in the calculation of the 4PI effective action, the only change in the expression in (\ref{gammaGen}) is that the non-interacting propagator is shifted
\bea
\label{shiftG}
i G_{\rm no\cdot int} ~~ \to ~~i G_{\rm no\cdot int\cdot\kappa} = i G_{\rm no\cdot int\cdot\kappa} - \hat R_\kappa\,.
\eea
The last step is to define an effective action that corresponds to the original classical action at the ultraviolet scale $\mu$ by making an additional shift to obtain
\bea
\label{hatGamma}
\Gamma_\kappa = \hat\Gamma_\kappa -\Delta S_\kappa(\phi)\,.
\eea
We use the notation $\Gamma = -i \Phi$ and we define an imaginary regulator function $R_\kappa = -i \hat R_\kappa$ (the extra factor $i$ will be removed when we rotate to Euclidean space). 
Using this notation equations (\ref{gammaGen}, \ref{shiftG}) become
\bea
\label{phi-def}
&& G_{\mathrm{no}\cdot \mathrm{int}\cdot\kappa}^{-1} = G_{\mathrm{no}\cdot \mathrm{int}}^{-1}-R_\kappa \,\\ [4mm]
&& \Phi_\kappa = 
-\frac{1}{2}{\rm Tr}\,{\rm ln} G^{-1} - \frac{1}{2} G_{\mathrm{no}\cdot \mathrm{int}\cdot\kappa}^{-1}G
+ \Phi_{{\rm int}}\,.
\label{phi-def2}
\eea

\section{Method}
\label{section-method}

For any value of $\kappa$ we define the functions $G_\kappa$ and $V_\kappa$ to be the self-consistent solutions that minimize the action. 
These solutions are not determined directly, but instead are obtained by solving a set of coupled differential flow equations. There are several steps involved in the derivation of these equations, which we explain in this section. 

We define kernels by taking functional derivatives of the action:
\bea
\Lambda^{(m,n)} = 2^m 4!^n G^{-4n} \frac{\delta}{\delta G^m}\frac{\delta}{\delta V^n} \Phi_{\rm int}[G,V]\,.
\eea
Subsitituting the (as yet unknown) self-consistent solutions we obtain $\kappa$ dependent kernels 
\bea
\label{kernels-RG-kappa}
&& \Lambda_\kappa^{(m,n)} = \Lambda^{(m,n)} \bigg|_{\stackrel{G=G_\kappa}{V=V_\kappa}}\,.
\eea
These kernels satisfy flow equations which can be found by a simple application of the chain rule
\bea
\label{flow1}
\partial_\kappa \Lambda_\kappa^{(m,n)} = \frac{\partial G_\kappa}{\partial \kappa} \frac{\delta}{\delta G_\kappa} \Lambda_\kappa^{(m,n)} +
\frac{\partial V_\kappa}{\partial \kappa} \frac{\delta}{\delta V_\kappa} \Lambda_\kappa^{(m,n)}\,.
\eea

\begin{figure}[h]
\begin{center}
\includegraphics[width=14cm]{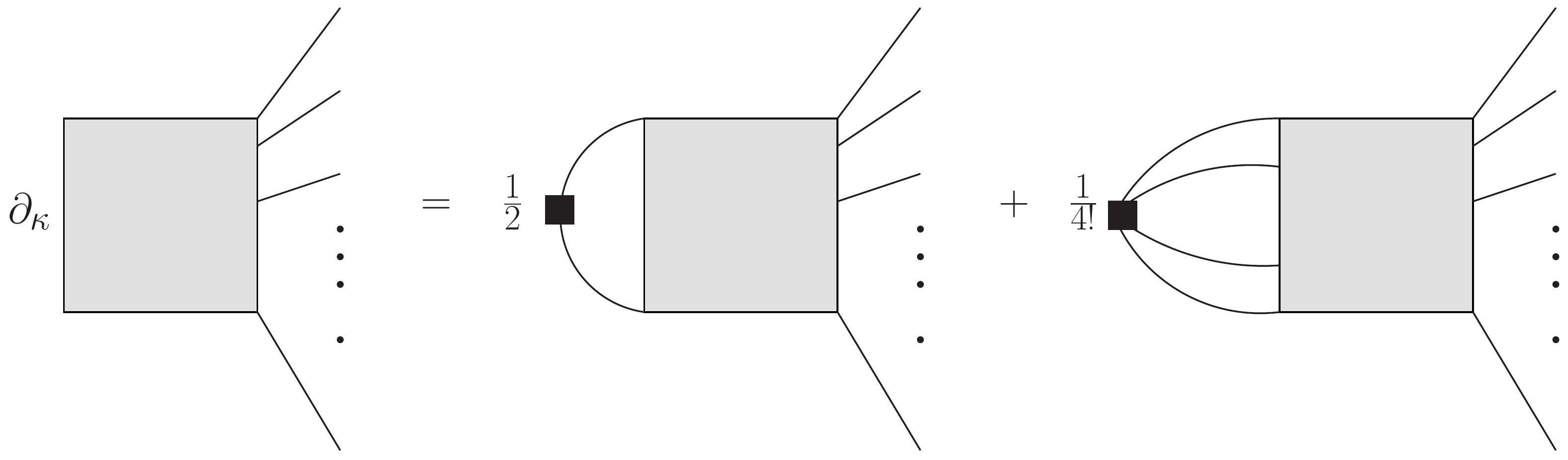}
\end{center}
\caption{Representation of equation \ref{flow2}. 
The black box on the line in the first figure on the right side represents the insertion $\partial_\kappa(\Sigma_\kappa+R_\kappa)$ and the black box on the vertex in the second figure represents $\partial_\kappa V_\kappa$. The vertical dots indicate that there are in total $2m+4n$ legs on the right side of each kernel. \label{flow-fig}}
\end{figure}

We can rewrite (\ref{flow1}) in a more convenient form. 
First we choose special names for three kernels we will write repeatedly  
\bea
\label{kernels-specific2}
\Lambda_\kappa^{(1,0)} = \Sigma_\kappa \,,~~~ \Lambda_\kappa^{(2,0)} = \Lambda_\kappa \,,~~~ \Lambda_\kappa^{(0,1)} = \chi_\kappa  \,.
\eea
Second we use the stationary conditions to rewrite the derivatives of the variational functions. 
The stationary condition for the 2 point function is 
\bea
\label{stat-cond}
\frac{\delta \Phi_\kappa[G,V]}{\delta G}\bigg|_{\stackrel{G=G_\kappa}{V=V_\kappa}} = 0
\eea
and using equations (\ref{phi-def}, \ref{phi-def2}, \ref{kernels-specific2}) we obtain
\bea
\label{dyson1-RG}
&& G_\kappa^{-1}  =  G_{\mathrm{no}\cdot \mathrm{int}}^{-1} - R_\kappa - \Sigma_\kappa\, \\
\label{eq1a}
&& \partial_\kappa G_\kappa = -G_\kappa\,(\partial_\kappa G^{-1}_\kappa)\, G_\kappa
 = G_\kappa\,\big(\partial_\kappa(R_\kappa+\Sigma_\kappa)\big)\,G_\kappa\,.
\eea
The stationary condition for the 4 point function can be written
\bea
\label{stat-cond2}
\frac{\delta \Phi_\kappa[G,V]}{\delta V}\bigg|_{\stackrel{G=G_\kappa}{V=V_\kappa}} = 0
\eea
and using equations (\ref{phi-def2}, \ref{kernels-specific2}) and Fig. \ref{phi-fig} we find
\bea
\label{partialV}
V_\kappa = V_0 + \chi_\kappa ~~~\to~~~ \partial_\kappa V_\kappa = \partial_\kappa \chi_\kappa\,.
\eea
Using equations (\ref{kernels-RG-kappa}, \ref{dyson1-RG}, \ref{partialV}) to rewrite (\ref{flow1}) we obtain
\bea
\label{flow2}
\partial_\kappa \Lambda_\kappa^{(m,n)} = \frac{1}{2} \partial_\kappa[\Sigma_\kappa +R_\kappa]G_\kappa^2 \Lambda_\kappa^{(m+1,n)}
+ \frac{1}{4!}\partial_\kappa V_\kappa G_\kappa^4 \Lambda_\kappa^{(m,n+1)}\,.
\eea
Equation (\ref{flow2}) gives a coupled hierarchy of integral equations. 
They are shown diagramatically in Fig. \ref{flow-fig}.
\noindent Continuum non-perturbative methods typically produce  hierarchies of integral equations and, as stated previously, a key feature of the $n$PI formalism is that the hierarchy in (\ref{flow2}) truncates naturally when the action is truncated. 
This point is difficult to understand without seeing the detailed structure of the kernels themselves, and therefore we will go ahead and write down the flow equations that we will solve, and then explain how the truncation works. 

\begin{figure}[h]
\begin{center}
\includegraphics[width=14cm]{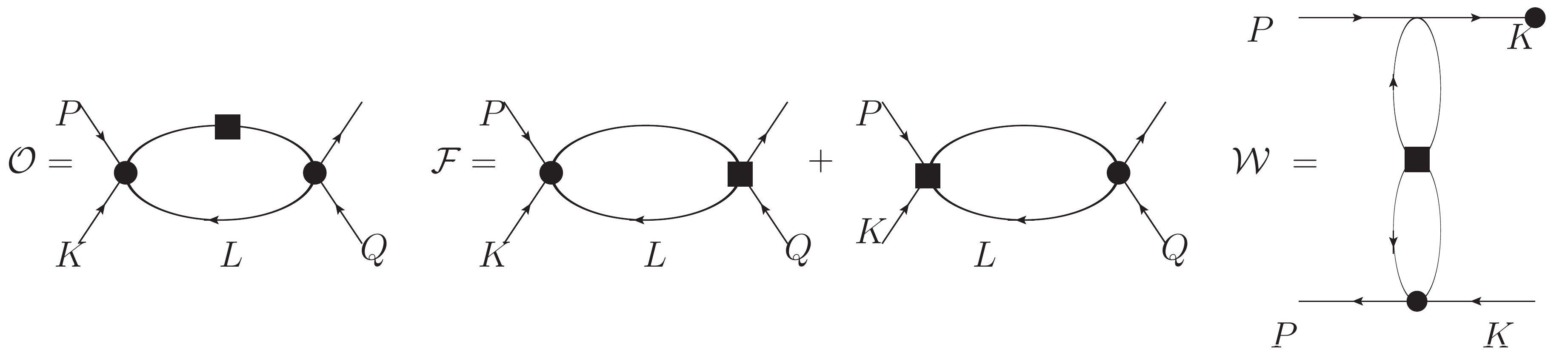}
\end{center}
\caption{Representation of the functions in equations (\ref{owl-def-4pi}, \ref{finch-def}, \ref{wren-def}). \label{owl-finch-wren-fig}}
\end{figure}

We Fourier transform and write the flow equations (\ref{flow2}) in  momentum space. 
We will also rotate to Euclidean space and use from now on only  Euclidean variables, without introducing addition subscripts to distinguish Euclidean quantities (details on our sign conventions are given in Ref. \cite{sohrabi}).
%
We will solve three flow equations, for the three kernels in (\ref{kernels-specific2}), which are obtained using 
$(m,n)=(1,0)$, $(2,0)$ and $(0,1)$ on the left side of (\ref{flow2}).
The basic structure of these equations in momentum space is
\bea
 \partial_\kappa \Sigma &=& \int \Lambda \,\big[~~~\big] + \int \Lambda^{(1,1)} \,\big[~~~\big] \nonumber\\
\partial_\kappa \Lambda &=& \int \Lambda^{(3,0)} \,\big[~~~\big] + \int \Lambda^{(2,1)} \,\big[~~~\big] \nonumber\\
\partial_\kappa V &=& \int \Lambda^{(1,1)}\,\big[~~~\big] + \int \Lambda^{(0,2)} \,\big[~~~\big] \,.
\nonumber
\eea
The kernels $\Lambda^{(1,1)}$ and $\Lambda^{(3,0)}$ and $\Lambda^{(0,2)}$ on the  right sides are calculated from their definitions (\ref{kernels-RG-kappa}) by taking functional derivatives of the effective action, and substituted into the integral equations. This procedure produces a closed set of equations - this is the truncation. 
The numerical method involves defining three functions, which are then interpolated inside the integrands of the three flow equations. These three functions are
\bea
\label{owl-def-4pi}
&& {\cal O}(P,K,Q) = \int dL\;  V(P,K,L) \;G_\kappa^2(L) \partial_\kappa\big[\Sigma_\kappa(L)+R_\kappa(L)\big] \; G(P+K+L) \; V(Q,-L,-P-K-Q) \nonumber \\ \\
\label{finch-def}
&& {\cal F}(P,K,Q) = \int dL\; \big[ \partial_\kappa V_\kappa(P,K,L) \;G_\kappa(L) \; G_\kappa(P+K+L) \; V_\kappa(Q,-L,-P-K-Q) \nonumber\\ [2mm]
 &&  ~~~~~~~~~~~~~~~~~~~~  ~~~~~ +  V_\kappa(P,K,L) \;G_\kappa(L) 
 \; G_\kappa(P+K+L) \; \partial_\kappa V_\kappa(Q,-L,-P-K-Q)\big] \nonumber \\ \\
&& {\cal W}(P,K) = \int dQ \int dL \nonumber \\
&& V_\kappa(P,-K,L)G_\kappa(L)G_\kappa(P-K+L)\,\partial_\kappa V_\kappa(Q,-L,P-K+Q)\,G_\kappa(Q) G_\kappa(P-K+Q) V_\kappa(-P,K,-Q)\,,\nonumber \\ \label{wren-def}
\eea
where the euclidean inverse propagator is
\bea
\label{dyson1-RG-euc}
&& G_\kappa^{-1}  =  G_{\mathrm{no}\cdot \mathrm{int}}^{-1} + R_\kappa + \Sigma_\kappa\,.
\eea
In Fig. \ref{owl-finch-wren-fig} we give a diagrammatic representation of these equations. 

\noindent In terms of these functions, the three flow equations are:
\bea
\label{flow-sigma2}
&& \partial_\kappa\Sigma_\kappa(P) = \frac{1}{2}\int dK \,\partial_\kappa\big[\Sigma_\kappa(K) + R_\kappa(K)\big]G_\kappa^2(K)\Lambda_\kappa(P,K)  -\frac{1}{4}\int dK {\cal W}(P,K) G_\kappa(K)  \nonumber \\ \\[.2cm]
&& \partial_\kappa \Lambda_\kappa(P,K)  = \partial_\kappa \bar\Lambda_\kappa(P,K) + \partial_\kappa \bar\Lambda_\kappa(P,-K) \nonumber \\
&& \partial_\kappa \bar \Lambda_\kappa(P,K)  = {\cal O}(P,K,-P)   + \frac{1}{4}{\cal W}(P,K) \nonumber \\
&& ~~~~~~~~~~~~~~~ +  \int dQ \;G_\kappa(Q) \, G_\kappa(P+K+Q)V_\kappa(P,K,Q) \bigl[ 2{\cal O}(P,Q,K) +  {\cal F}(P,Q,K)\bigr]   \nonumber \\
\label{newLambda}\\
&& \partial_\kappa V_\kappa(P,K,Q) = {\cal O}(P,K,Q)+{\cal O}(K,Q,P)+{\cal O}(Q,P,K) \nonumber \\ 
&& ~~~~~~~~~~~~~~~~~~ + 
\frac{1}{2}\big[{\cal F}(P,K,Q)+{\cal F}(K,Q,P)+{\cal F}(Q,P,K)\big]\,.
\label{flowV}
\eea

To solve these differential equations, we need to define boundary conditions. To see how to do this, we recall the discussion about the definition of the regulator function $R_\kappa(Q)$. This function is chosen so that at the ultraviolet scale $\kappa=\mu$ all fluctuations are suppressed and the theory is essentially classical, while in the limit $\kappa\to 0$ the regulator disappears and the full quantum theory is restored. 
The strategy is therefore to solve the flow equations starting from the scale $\kappa=\mu$ and using the known classical solutions as boundary conditions, and then extract at the $\kappa=0$ end of the flow the quantum $n$-point functions that we are looking for. 

The regulator function we use has the form \cite{wetterich}
\bea
\label{Rdef}
R_\kappa(Q) = \frac{Q^2}{e^{Q^2/\kappa^2}-1}\,
\eea
and our initial conditions for the flow equations are
\bea
&& G_\mu^{-1} = P^2+m^2 +\Sigma_\mu(P)  \text{~~~ and~~~} \Sigma_\mu(P) = m_b^2-m^2 \nonumber \\
\to ~~ && G_\mu^{-1} = P^2+m_b^2 \\
&& \Lambda_\mu(P,K) = V_\mu(P,K,Q) = -\lambda_b
\eea
where $m_b$ and $\lambda_b$ are the bare parameters of the original Lagrangian. 
The physical parameters are defined through the renormalization conditions
\bea
\label{renorm-cond}
G_0^{-1}(0) = m^2 \text{~~~ and~~~} V_0(0) = -\lambda \,.
\eea

The goal of the renormalization program is to absorb all divergences into the definitions of the bare parameters. This is possible if the truncation is performed correctly. In practical terms, it means that the equations we solve must not contain any unregulated loops in the limit $\kappa\to 0$. 
To see how this condition is satisfied for the three functions $({\cal O}, {\cal F}, {\cal W})$ we look at Fig. \ref{owl-finch-wren-fig}.
Each diagram in the figure contains a loop that would be logarithmically divergent, if it were not for the presence of the regulator 
(recall that a black box on a line indicates the insertion $\partial_\kappa(\Sigma_\kappa+R_\kappa)$ and a black box on a vertex represents $\partial_\kappa V_\kappa$, and both of these quantities go to zero in the limit $\kappa\to 0$ where the regulator disappears). 

\begin{figure}[h]
\begin{center}
\includegraphics[width=14cm]{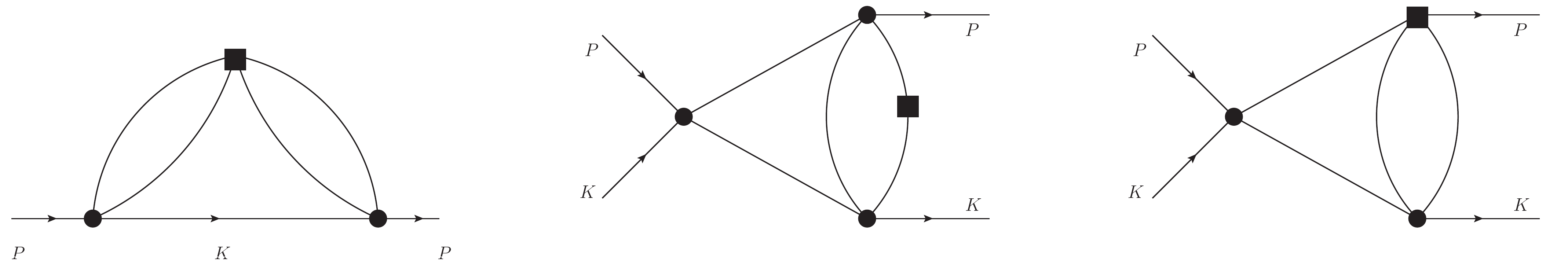}
\end{center}
\caption{The first diagram represents the integral in the second term of equation (\ref{flow-sigma2}), and the second and third diagrams are respectively the first and second integrals in equation (\ref{newLambda}). \label{flow-ints-fig}}
\end{figure}

We must also consider the loop structures that appear when the functions in equations (\ref{owl-def-4pi}, \ref{finch-def}, \ref{wren-def}) are embedded into the kernels of the flow equations  (\ref{flow-sigma2}, \ref{newLambda}, \ref{flowV}). 
A diagrammatic representation of 
the integral in the second term of (\ref{flow-sigma2}) and the two contributions to the integral in the third term of (\ref{newLambda}) are shown in Fig. \ref{flow-ints-fig}. 
The figure shows that the loops that are formed by joining legs of the functions $({\cal O}, {\cal F}, {\cal W})$ are not divergent. 
The integral in the first term on the right side of (\ref{flow-sigma2})  is represented by the first diagram on the right side of Fig. \ref{flow-fig}. In this graph, the grey box is the kernel $\Lambda$ which is rendered finite by its own flow equation, and the loop that is formed by joining two legs of the 4-kernel is regulated by the insertion $\partial_\kappa(\Sigma_\kappa+R_\kappa)$ that is represented by the small black box. 

Now we return to the issue of the truncation. We stated earlier that it will be possible to absorb all divergences into the definitions of the bare parameters if the truncation is done correctly, and that a correct truncation is one that ensures that all potentially divergent loops contain an insertion that goes to zero in the limit $\kappa\to 0$. 
To understand this better, we consider what would happen if we ignored the flow equation for the  kernel $\Lambda$. We could calculate the kernel $\Lambda$ from its definition (\ref{kernels-RG-kappa}) and substitute it directly into the $\Sigma$ flow equation (\ref{flow-sigma2}), 
which would eliminate the need to solve the $\Lambda$ flow equation (\ref{newLambda}). 
However, if we do this, the right side of the $\Sigma$ flow equation will contain 2 loop diagrams with unregulated sub-divergences (one example is the first diagram in Fig. \ref{flow-ints-fig} but with the insertion $\partial_\kappa V_\kappa$  - which is represented by the black box - 
replaced with the normal vertex $V_\kappa$).
This tells us that the kernel $\Lambda$ cannot be calculated directly but must be flowed. 
If we were to work beyond the four loop level then the kernels $\Lambda^{(1,1)}$ and $\Lambda^{(3,0)}$ and $\Lambda^{(0,2)}$ which were substituted directly to obtain our flow equations for the kernels $\Sigma=\Lambda^{(1,0)}$ and $\Lambda=\Lambda^{(2,0)}$, would themselves have to be flowed.


Finally we comment on the fact 
our method requires that we choose specific values of the bare parameters from which to start the flow. 
A different choice of bare parameters will  give different quantum $n$-point functions at the end of the flow, and therefore different renormalized masses and couplings. 
The procedure to figure out the values of the bare parameters that will satisfy the chosen renormalization conditions is called tuning. 
Starting from an initial guess for the bare parameters, we solve the flow equations, extract the renormalized parameters, adjust the bare parameters either up or down depending on the result, and solve the flow equations again. 
The calculation is repeated until the chosen renormalization conditions are satisfied to the desired accuracy. 

\section{Numerical Results}
\label{section-results}

Our renormalization conditions are defined in equation (\ref{renorm-cond}). 
We set $m=1$, and measure all dimensionful quantities in units of the mass. 
We choose $\lambda=2$. 

The differential equations are solved using a logarithmic scale by defining the variable $t=\ln \kappa/\mu$, in order to increase sensitivity to the small $\kappa$ region where we approach the quantum theory. We use $\kappa_{\rm max} = \mu=100$, $\kappa_{\rm min}=10^{-2}$ and $N_\kappa=50$. We have checked that our results are insensitive to these choices. 
The 4-dimensional momentum integrals are written in the imaginary time formalism as
\bea
\label{4dint}
\int dK \,f(k_0,\vec k) = T \sum_n \int\frac{d^3k}{(2\pi)^3}\,f(m_t n,\vec k)\,
\eea 
with $m_t = 2\pi T$. 
Numerically we take $N_t$ terms in the summation with $\beta = \frac{1}{T} = N_t a_t$ where the parameter $a_t=1/10$ is the lattice spacing in the temporal direction. 
%
The integrals over the 3-momenta are done in spherical coordinates and using Gauss-Legendre integration. 
The dependence on the angles is weak and results are very stable when the number of lattice points for the polar and azimuthal angles equals 4 or 6. 
To calculate the integral over the magnitude of the 3-momenta, we define a spatial length scale analogous to the inverse temperature $L=a_s N_s$ where $a_s$ is the spatial lattice spacing and $N_s$ is the number of lattice points. In momentum space we have $p_{\rm max} = \pi/a_s$ and the momentum step is characterized by the parameter $\Delta p = \pi/(a_s N_s)$.

\begin{figure}[h]
\begin{center}
\includegraphics[width=10cm]{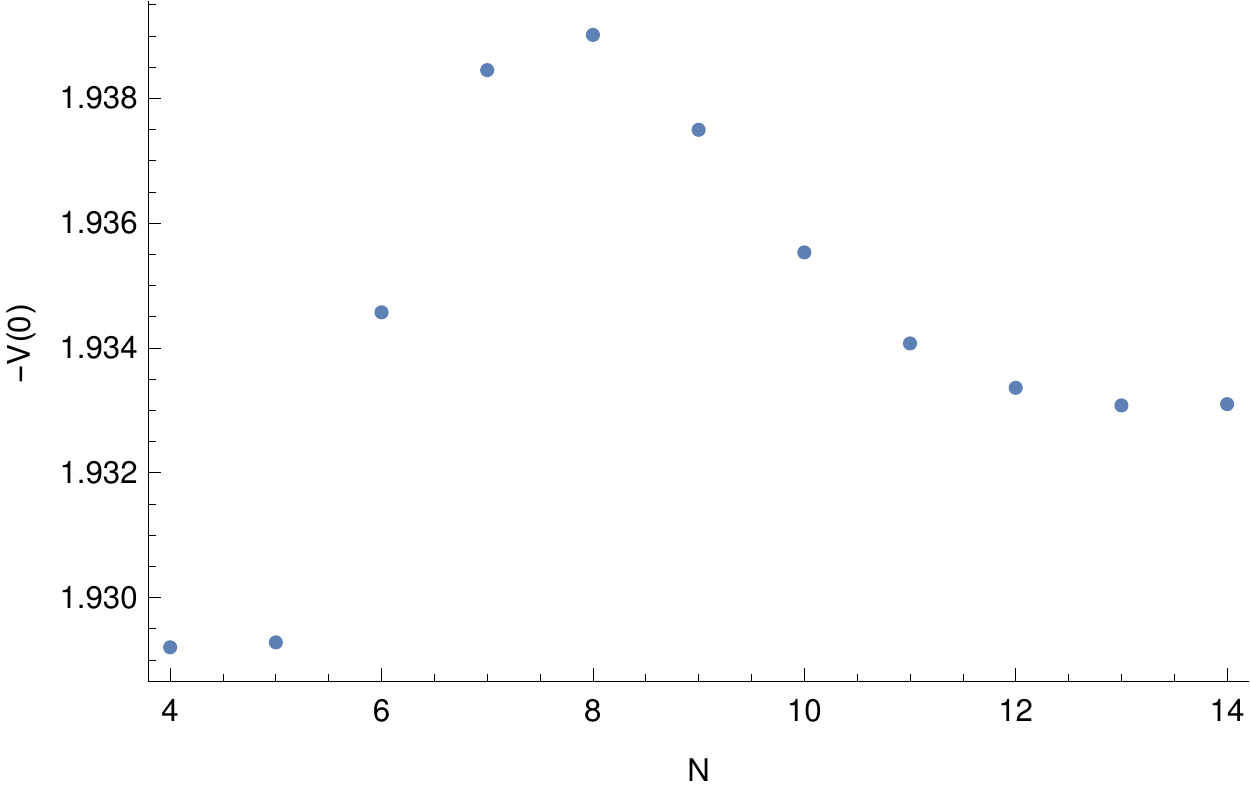}
\end{center}
\caption{The vertex at zero momentum as a function of the number of spatial lattice points $N_s$ with fixed $a_s=1/8$ (or $p_{\rm max}$ =25.1). \label{convergence-fig}}
\end{figure}

In Fig. \ref{convergence-fig} we show a plot of the vertex at zero momentum versus the number of spatial grid points $N_s$ with $a_s$ held fixed at 1/8 (which corresponds to holding $p_{\rm max}$ fixed while decreasing $\Delta p$). The figure shows that convergence is achieved with $N\approx 12$. 

\begin{figure}[htbp]
\begin{center}
\includegraphics[width=8cm]{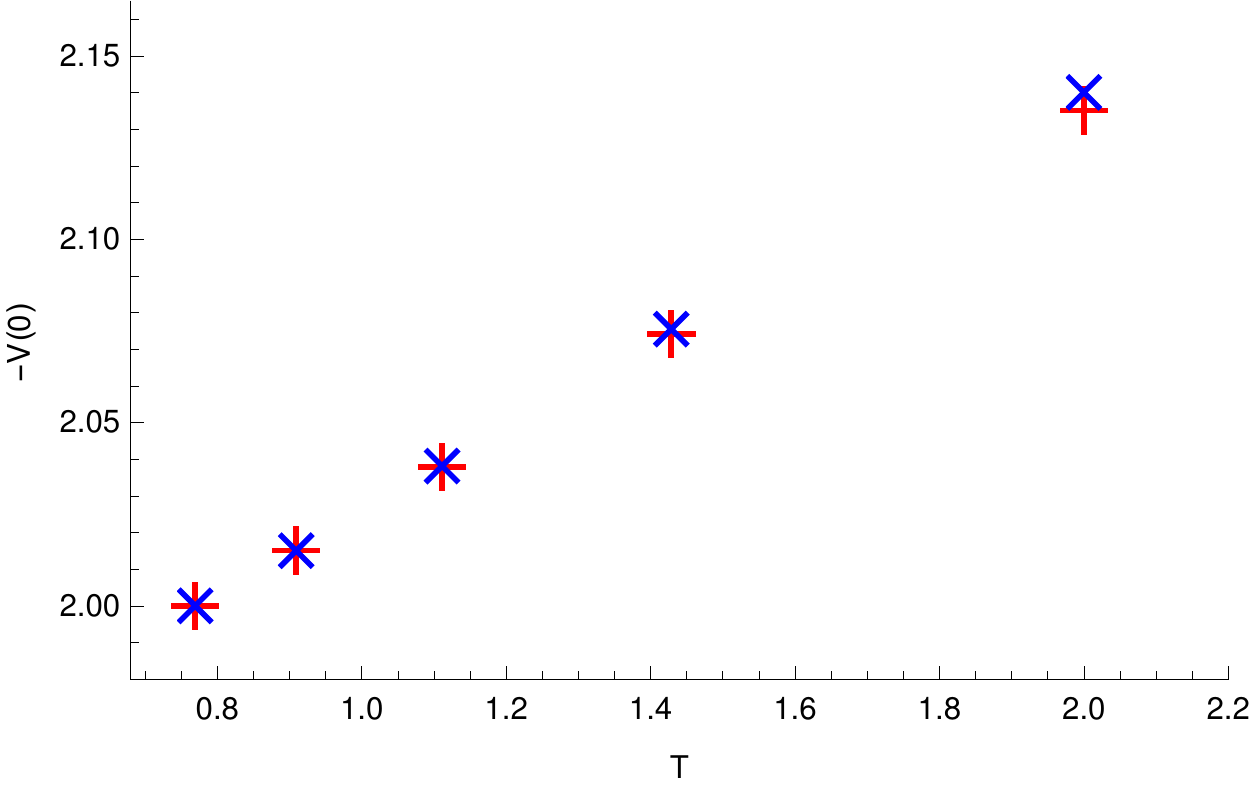}
\end{center}
\caption{
The vertex at zero momentum as a function of temperature using two different renormalization points: $T_{\rm renorm}$=2.0 (red +) and $T_{\rm renorm}$=0.91 (blue x).\label{scale-fig}}
\end{figure}

Ideally we want to renormalize at a temperature close to zero, but smaller temperatures correspond to larger values of $N_t$ and therefore to increase the speed of the calculation we would like to use a higher temperature. To check that this is okay, we test the scale dependence of our results. In Fig. \ref{scale-fig} we show the vertex at zero momentum as a function of temperature, with the renormalization done at two different temperatures. To make a physical comparison the curves are shifted so that they agree at the smallest temperature, and both satisfy the same renormalization condtion
$V_0(0)|_{T=T_{\rm min}}=-\lambda$ [see equation (\ref{renorm-cond})]. The slight shift of the curves relative to each other at the highest temperature is a measure of the scale dependence of the calculation. 

\begin{figure}[!htbp]
\begin{center}
\includegraphics[width=8cm]{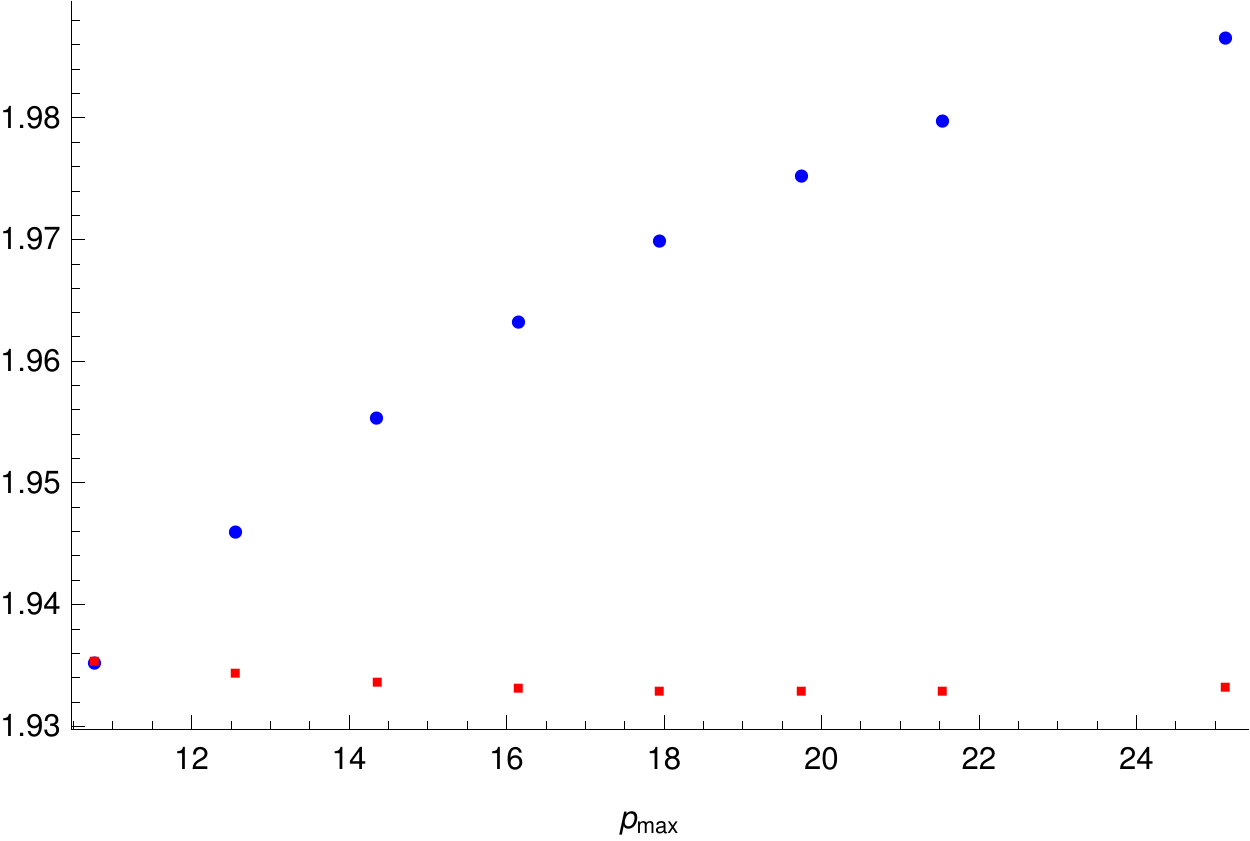}
\end{center}
\caption{
The vertex at zero momentum (red squares) and bare coupling (blue dots) as a function of $p_{\rm max}$ with $\Delta p$ held fixed, using $T=1.4$ and $L=14/8$.
\label{renorm-fig}}
\end{figure}

The main result of this paper is to show that using the FRG renormalization method that we have introduced, all divergences are absorbed by the bare parameters of the Lagrangian. 
In Fig. \ref{renorm-fig} we show the bare vertex and the self-consistent vertex at zero momentum as functions of the momentum cutoff, with $T$ and $\Delta p$ held fixed. 
%
The figure shows clearly that as the momentum cutoff increases, the divergence in the vertex is absorbed by the bare coupling, while the renormalized vertex remains approximately constant. This result proves that our renormalization method is working correctly. 
\clearpage

\section{Conclusions}
\label{section-conclusions}


The 4PI equations of motion involve
nested subdivergences that cannot be cancelled using a
finite number of counterterms. In this paper we have
introduced a completely new method to renormalize the
4PI theory which is based on a renormalization group
approach. 
Calculations involving vertex corrections are computationally intensive, but we have obtained preliminary results that prove the success of our method. 
Using a symmetric scalar $\phi^4$ theory at the 4 loop 4PI level, we have shown that all divergences are absorbed into the bare parameters of the Lagrangian, and a finite physical mass and coupling are obtained.
Using this method makes is possible, for the
first time, to use $n$PI effective action techniques beyond the 2PI level.
The next step will be to calculate the thermodynamic pressure in the same scalar theory and compare with the results of  perturbative and 2PI calculations. This calculation is in progress.

\end{document}